\documentclass[aps,nofootinbib,showpacs,showkeys]{revtex4}
\usepackage{graphicx,amsmath}
\begin{document}

\title{A note on Charmed and Bottomed Pentaquark Production by Fragmentation}
\author{Kingman Cheung}
\affiliation{Department of Physics and NCTS, National Tsing Hua University,
Hsinchu, Taiwan R.O.C.}
\date{\today}

\begin{abstract}
H1 Collaboration recently observed the charmed pentaquark.  
In this short note, we point out that the dominant production mechanism 
for pentaquark consisting of a heavy quark is heavy quark fragmentation.
We obtain a crude estimate on the fragmentation probability for charm
quark into $\Theta_c^0$, based on the known fragmentation probabilities of
charm quark into mesons and baryons:  $f(\bar c \to \Theta_c^0)
\simeq (2-7) \times 10^{-3}$.  Similarly, we also obtain the fragmentation
probability for bottom quark into $\Theta_b^+$: 
$f(\bar b \to \Theta_b^+) \simeq (5-20) \times 10^{-3}$.
We also estimate the prospect of observing $\Theta_c^0$ and $\Theta_b^+$
at HERA, LEP, and Tevatron.
\end{abstract}
\preprint{}
\maketitle

\section{Introduction}
H1 Collaboration recently reported the discovery of a charmed pentaquark
$\Theta_c^0$ \cite{H1,nopenta}.  This is the first evidence of pentaquark 
consisting of a heavy (anti-)quark, following the excitement 
of $\Theta^+(1540)$ in 2003 \cite{exp1}.  
The interpretation of the observed charmed pentaquark follows closely
along with the $\Theta^+$, with the $\bar s$ replaced by $\bar c$.

In the constituent quark model, it is believed that the $\Theta^+ (1540)$ 
consists of $\bar s uudd$.  There are various possible configurations 
for this complicated
system.  Jaffe and Wilczek (JW) \cite{wilczek} interpreted the bound state
as a diquark-diquark-antiquark structure.  Each diquark
pair is in the $\bar{\textbf{3}}_c$ representation of SU(3)$_c$ (analogous
to an antiquark) and in the spin state $S=0$.  Thus,
the two diquark pairs combine in a $P$-wave orbital angular momentum
to form a state with $\textbf{3}_c$ in color, 
spin $S=0$, and $\bar{\textbf{6}}_f$ in flavor.  Then, combining with
the antiquark to form a flavor antidecuplet and octet with spin $S=1/2$.
The $\Theta^+$ is at the top of the antidecuplet and has an isospin $I=0$.
On the other hand, Karliner and Lipkin (KL) \cite{lipkin} interpreted
the bound state as a diquark-triquark $(ud)$-$(ud\bar s)$ structure.  
The first 
stand-alone $(ud)$ diquark pair is in a state of spin $S=0$, color 
$\bar{\textbf{3}}_c$ and flavor $\bar{\textbf{3}}_f$  while the second
$(ud)$ diquark pair inside the cluster $(ud\bar s)$ is in a state of
spin $S=1$, color $\textbf{6}_c$ and flavor $\bar{\textbf{3}}_f$.
The triquark cluster is then in a state of spin $S=1/2$, color
$\textbf{3}_c$ and flavor $\bar{\textbf{6}}_f$.  So the overall configuration
will give a color singlet, spin $S=1/2$, and a flavor octet and antidecuplet.
This internal configuration of KL is different from that of JW,
as the diquark pairs in the JW configuration are symmetric while those 
pairs in the LK configuration are asymmetric.
The present author \cite{cheung} examined these two internal structures
using color-spin hyperfine interactions, and found that the 
diquark-diquark-antiquark structure is slightly favorable in terms of
hyperfine interaction.  However, there could also be  a mixing 
between these two configurations.

The diquark-diquark-antiquark picture of Jaffe and Wilczek \cite{wilczek}
can be easily extended to charmed pentaquark, with the replacement
$\bar s \to \bar c$.  Since the charm quark does not belong to the
SU(3)$_f$ of $(u,d,s)$, the internal quark configuration of $\Theta_c^0$ will
follow the configuration of the diquark-diquark subsystem.
The masses of the charmed and bottomed pentaquark were also estimated using 
various methods \cite{wilczek,sasaki,lipkin,cheung,chiu}.  
Jaffe and Wilczek \cite{wilczek} estimated the mass of $\Theta_c^0$ 
to be $2710$ MeV, Sasaki \cite{sasaki} using lattice QCD 
estimated the $\Theta_c^0$ to be about 640 MeV above the $DN$ threshold,
Karliner and Lipkin \cite{lipkin} gave a value of 
$2985 \pm 50$ MeV, Cheung \cite{cheung} estimated the mass to be
between 2938 and 2997 MeV, and Chiu and Hsieh using the lattice gauge 
method \cite{chiu} obtained a value of $2977(109)$ MeV.
\footnote
{Stewart, Wessling, and Wise \cite{wise} had estimates on the mass of the 
S-wave charmed pentaquark $T_s$ and the S-wave bottomed pentaquark $R_s$,
which are below the strong decay thresholds.
}
The latter three
estimations are about 100 MeV, which is of the order of the uncertainty
in the method of estimation,  below the experimental value of 3099 MeV
\cite{H1}.

Unfortunately, another collaboration, ZEUS, at HERA so far has not reported
any positive evidence for the charmed pentaquark \cite{nopenta}.  
While we are still waiting
for further data, it is timely to point out that the 
dominant production mechanism for pentaquarks consisting of a heavy
(anti-)quark is heavy quark fragmentation, similar to the production of
heavy-light mesons and baryons consisting of a heavy quark.
In this note, we obtain an estimate on the fragmentation
probability for a heavy quark into a pentaquark.  Using the probabilities
we can then estimate the production rates at various collider environments,
like HERA, LEP, and Tevatron. 

The organization is as follows.  In the next section, we estimate the
fragmentation probabilities.  In Sec. III, we estimate the production rates
of $\Theta_c^0$ and $\Theta_b^+$ at HERA, LEP, and Tevatron.  We summarize
in Sec. IV.

\section{Estimates of fragmentation probabilities}
A naive estimate of the $\Theta_c^0$ mass is as follows. The difference between
$\Lambda_c$ and a constituent charm quark (one half of $J/\psi$) is
that $\Lambda_c$ has an additional $ud$ diquark.  The constituent mass
of this diquark is $M_{(ud)} \equiv M_{\Lambda_c} - M_{\psi}/2 = 736.47$ MeV.
The constituent content of $\Theta_c^0$ has one more diquark
than $\Lambda_c$, and thus the mass of $\Theta_c^0$ is 
$M_{\Theta_c^0}= M_{\Lambda_c} + M_{(ud)} = 3021.4$ MeV, which is 
amazingly close to previous estimates \cite{lipkin,cheung,chiu}
and the observed value by H1 \cite{H1}.  
We can use the same naive method to obtain the estimate for the mass of
$\Theta_b^+$: $M_{\Theta_b^+} = M_{\Lambda_b} + (M_{\Lambda_b} - M_{\Upsilon}
/2) = 6518$ MeV, which is within 100 MeV uncertainty of our previous
estimate based on color-spin hyperfine interaction \cite{cheung}. 

Such an agreement between the naive estimate and the observed value may 
suggest that the dominant production mechanism for $\Theta_c^0$
and $\Theta_b^+$ is the same as $\Lambda_c^+$ and $\Lambda_b^0$, respectively,
i.e., by heavy quark fragmentation.  Let us first focus on estimating
the fragmentation probability $f(\bar c \to \Theta_c^0)$, then repeat the
exercise for $f(\bar b \to \Theta_b^+)$.  
The fragmentation probability for the charm quark into $D$ mesons, summing
over $D^0$ and $D^+$ is \cite{cfrag}
\[
f\left (c \to (c\bar u), (c\bar d) \right )\simeq 0.781 \pm 0.023 \;,
\]
which implies the average 
\[
f(c \to c\bar q) \simeq 0.39 \pm 0.012
\]
for one flavor of light quark $q$.
We can then compare the probability into a diquark $(dq)$
and a single quark as
\[
\frac{ f(\bar c \to \bar c (dq) )}{ f(\bar c \to \bar c q )} =
\frac{ f\left( c \to c(ud) \right)}{ f(c \to c \bar q) )} 
= \frac{0.076 \pm 0.007}{0.39 \pm 0.012 } 
= 0.195 \pm 0.019 \;,
\]
where we have used  $f(c \to c (ud))= f(c\to \Lambda_c^+)=0.076\pm 0.007$.  
Therefore, the fragmentation probability for $\bar c \to \Theta_c^0$ is
\begin{eqnarray}
f \left(\bar c \to \bar c (ud)(ud) \right ) &=& 
f \left(\bar c \to \bar c (dq)(dq) \right) \nonumber
\\
& = & f (\bar c \to \bar c \bar q \bar q  ) \times 
\frac{ f(\bar c \to \bar c (dq) )}{ f(\bar c \to \bar c q )} \times
\frac{ f(\bar c \to \bar c (dq) )}{ f(\bar c \to \bar c q )} \nonumber \\
&=& (2.89 \pm 0.82) \times 10^{-3} \;.
\end{eqnarray}
This is the prediction by the method I.
An alternative way of thinking (method II) is a two-step fragmentation process:
\begin{equation}
f (\bar c \to \bar c (ud)(ud) ) = f (\bar c \to \bar c u d) \times
 f ( (\bar c ud) \to (\bar c u d) u d) = (0.076 \pm 0.007)^2 = 
(5.78 \pm 1.06) \times 10^{-3} \;.
\end{equation}
These two ways of visualizing the fragmentation process give consistent
probabilities within a factor of two.  We therefore take the range to be
\begin{equation}
\label{thetaC}
f(\bar c \to \Theta_c^0) \simeq (2-7) \times 10^{-3}\;,
\end{equation}
which is obtained by considering $-1\sigma$ from the lower value and
$+1\sigma$ from the upper value.
In the H1 pentaquark paper \cite{H1}, they also 
mentioned that the fraction of $\Theta_c^0$ is roughly 1\% of the total
$D^*$ production, which means 
\begin{equation}
f_{\rm exp} (\bar c \to \Theta_c^0) \simeq
10^{-2} \times f(c\to D^*) = 2.35 \times 10^{-3} \;,
\end{equation}
where we have used $f(c\to D^*)= 0.235$ \cite{cfrag}.

We can now repeat the above exercise for the bottom quark. The method I gives
\begin{eqnarray}
f(\bar b \to \Theta_b^+) &=& f(\bar b \to \bar b (dq)(dq)  )\nonumber \\
&=&  f (\bar b \to \bar b \bar q \bar q  ) \times 
\frac{ f(\bar b \to \bar b (dq) )}{ f(\bar b \to \bar b q )} \times
\frac{ f(\bar b \to \bar b (dq) )}{ f(\bar b \to \bar b q )} \nonumber \\
&=& (1.09 \pm 0.56)  \times 10^{-2} \;,
\end{eqnarray}
where we have used $f(\bar b \to \bar b q)= 0.388\pm 0.013$, 
$f(b \to b{\rm -baryon}) =0.118\pm 0.02$ \cite{pdg}. 
Using the method II, we obtain
\begin{equation}
f (\bar b \to \Theta_b^+ ) = \left(f (b \to b{\rm -baryon}) \right)^2 =
(1.39 \pm 0.47) \times 10^{-2} \;.
\end{equation}
Thus, the estimated range is
\begin{equation}
\label{thetaB}
f( \bar b \to \Theta_b^+ ) = (5 - 20) \times 10^{-3}\;.
\end{equation}
Note that in our estimate $f(\bar b \to \Theta_b^+ ) >
f(\bar c \to \Theta_c^0 )$.  This is easy to understand, because the heavier 
$b$ quark can extract the light quarks from the vacuum easier than the charm
quark.  Thus, in this sense the chance of forming $\Theta_b^+$ is easier
than forming $\Theta_c^0$.

\section{Production rates}
Once we are equipped with the fragmentation probabilities we are ready to 
calculate the production rates of $\Theta_c^0$ and $\Theta_b^+$ at
various collider environments.  The cross section for $\Theta_c^0$ is 
given by
\begin{equation}
\sigma (\Theta_c^0) = \sigma( \bar c+X) \times f(\bar c \to \Theta_c^0) \;,
\end{equation}
and a similar formula for $\Theta_b^+$.

At the HERA $ep$ collider, the dominant production channel for heavy quarks
is $\gamma^* g \to Q \overline{Q}$ where $Q=c$ and $b$.  H1 Collaboration
has measured the open charm production \cite{H1-charm} in the kinematic 
region: $ 2< Q^2 < 100$ GeV$^2$, $0.02<y <0.7$, and with $1.5 < p_T (D^{*\pm })
<15$ GeV and $|\eta(D^{*\pm })|<1.5$, the cross section
$\sigma(D^{*\pm }) \simeq 8$ nb.  Then, we can obtain the inclusive
$\sigma( c+X$ and $\bar c +X)=$$\sigma(D^{*\pm })/f(c\to D^{*\pm})=$
$8\,{\rm nb}/0.235 = 34$ nb.  Thus, the cross section for $\Theta_c^0$ is
\begin{equation}
\sigma(\Theta_c^0) = 34 \,{\rm nb} \times (2-7)\cdot 10^{-3} 
= 68 - 240 \;{\rm pb} \;,
\end{equation}
using Eq. (\ref{thetaC}).  H1 \cite{H1} with a luminosity of $75$ pb$^{-1}$ 
observed $50.6\pm 11.2$ signal
events, from which the observed $D^* p$ resonance is estimated to contribute
roughly 1\% of the total $D^*$ production in the kinematic region studied.
The detection efficiency is then of the order of $O(1\%)$ or less.

We can then proceed to the open bottom quark production at HERA.  A recent
measurement by ZEUS \cite{ZEUS-beauty} 
gives $b\bar b$ production in the kinematic region:
$Q^2 >2$ GeV, $0.05<y<0.7$ with one of the $b$ quarks hadronizing into a 
jet and the other $B$ decaying into a muon:
\[
\sigma_{b\bar b} ( ep \to e b \bar b +X \to e j \mu +X) = 40.9 \pm 5.7 \;
^{+6.0}_{-4.4} \;{\rm pb} \;.
\]
Note that the branching ratio $B(B\to \mu \bar \nu_\mu X) = 
10.38$\% \cite{pdg},  and we
assume that hadronization into a jet is 100\%.    We can obtain the total 
$b\bar b$ cross section
\[
\sigma(ep \to e b\bar b + X) = 394 \; ^{+80}_{-69} \; {\rm pb}
\]
Thus, the production cross section for $\Theta_b^+$ is 
\begin{equation}
\sigma(\Theta_b^+) = 2-8\;{\rm pb} \;,
\end{equation}
using Eq. (\ref{thetaB}).  With a 
luminosity of $O(100)$ pb$^{-1}$ but an efficiency of the order of 1\% or less
(see above for $\Theta_c^0$ detection), the number of $\Theta_b^+$ that
would be observed is less than 10 events.   Therefore, the prospect of 
observing $\Theta_b^+$ at HERA is not good.

We now turn to LEP $Z$ data.  With a total more than 10 millions hadronic 
$Z$ decays collected by all 
four collaborations at LEP, the prospect of observing $\Theta_c^0$
and $\Theta_b^+$ is in fact quite feasible.  In addition, the number of raw
$b\bar b$ events is slightly more than $c \bar c$.  Using $10^7$ hadronic
$Z$ decays and $R_c = 0.1720 \pm 0.0030$ and $R_b = 0.21638 \pm 0.00066$
\cite{lepew}, the number of raw $\Theta_c^0$ and $\Theta_b^+$ is about
$3400-12000$ and $11000-43000$ events, respectively.  Therefore, even with
an efficiency less than 1\% the chance of observing 
$\Theta_c^0$ and $\Theta_b^+$
is very optimistic, especially $\Theta_b^+$. 

We can also estimate the production rates at the Tevatron. A recent published
result \cite{cdf-charm} measured the production rates of $D^0,\; D^{*+},\,
D^+$, and $D_s^+$ at $\sqrt{s}=1.96$ TeV:
\begin{eqnarray}
\sigma(D^0, p_T>6\,{\rm GeV},\; |y|<1) &=& 9.4 \; \mu{\rm b} 
\nonumber \\
\sigma(D^{*+}, p_T>6\,{\rm GeV},\; |y|<1) &=& 5.2 \pm 0.1 \pm 0.8 \; \mu{\rm b}
\nonumber \\
\sigma(D^+, p_T>6\,{\rm GeV},\; |y|<1) &=& 4.3 \pm 0.1 \pm 0.7 \; \mu{\rm b}
\nonumber \\
\sigma(D_s^+, p_T>8\,{\rm GeV},\; |y|<1) &=& 0.75 \pm 0.05 \pm 0.22 \; 
\mu{\rm b} \;, \nonumber 
\end{eqnarray}
where we do not give the error for the $D^0$ case because we do not 
understand how they summed the errors of all the $p_T$ bins.  We then use
the fragmentation probability of charm quark into $D^0, \, D^+$, and 
$D^{*+}$ to convert the above cross sections into the total charm quark 
cross section $\sigma (c+X) \sim 20\;\mu$b.   
Thus, the production cross section for $\Theta_c^0$ is in the range
\begin{equation}
\sigma (\Theta_c^0) =  
(4-14) \times 10^4 \;{\rm pb} \;.
\end{equation}
The expectation for a luminosity of $O(100)$ pb$^{-1}$ and an 
efficiency of the 
order of 1\% is in the range of $(4-14) \times 10^4$ events of $\Theta_c^0$.
Thus, 
there is a good chance of observing $\Theta_c^0$
 even the efficiency is further worsen.

We can also estimate the production rate for $\Theta_b^+$ at the Tevatron.
We found a recent published result on $B^+$ production at $\sqrt{s}=1.8$ TeV
 \cite{cdf-bottom}:
\[
\sigma(B^+, p_T > 6 \;{\rm GeV},\; |y|<1) = 3.6 \pm 0.4 \pm 0.4 \; \mu{\rm b}
\;,
\]
which can be converted to raw $\bar b+X$ quark cross section by dividing the
cross section with $f(\bar b \to B^+) =0.388 \pm 0.013$ \cite{pdg}, and we
obtain $\sigma(\bar b+X) = 9.3 \pm 1.5\; \mu$b.   Therefore, the estimated
range of $\Theta_b^+$ cross section is
\begin{equation}
\sigma(\Theta_b^+) =  
(4 - 20)\times 10^4 \; {\rm pb} \;.
\end{equation}
The expectation for a luminosity of $O(100)$ pb$^{-1}$ and an 
efficiency of the 
order of 1\% is then in the range $(4- 20) \times 10^4$ events of $\Theta_b^+$.
Again, the chance of observing $\Theta_b^+$ at the Tevatron is very good
 even the efficiency is further worsen.

\section{Conclusions}
In this note, we have pointed out that the dominant production mechanism 
for pentaquark consisting of a heavy quark is heavy quark fragmentation, similar
to heavy-light mesons and baryons consisting of a heavy quark.  Based on
the known measurements of the probabilities of charm quark or bottom quark
into mesons and baryons, we have estimated the fragmentation probabilities for 
$\bar c \to \Theta_c^0$ and $\bar b \to \Theta_b^+$, as given by
$f(\bar c \to \Theta_c^0)\simeq (2-7)\times 10^{-3}$ and 
$f(\bar b \to \Theta_b^+)\simeq (5-20)\times 10^{-3}$.
The $\Theta_c^0$ has been observed at HERA, but the prospect of observing 
$\Theta_b^+$ at HERA 
is not good.  On other hand, the predicted number of events for $\Theta_c^0$
and $\Theta_b^+$ at LEP and at the Tevatron should be large enough for them to
be discovered.  

\acknowledgments
The work was supported by the NSC with the grant number 
NSC-92-2112-M-007-053-.


\begin{thebibliography}{99}
\bibitem{H1} 
H1 Collaboration, hep-ex/0403017.

\bibitem{nopenta}
K. Lipka for the H1 and ZEUS Collaborations, 
 ``{\it Charm production in ep interaction at HERA and evidence for 
a narrow anti-charmed baryon state at H1}'', hep-ex/0405051.

\bibitem{exp1}
T. Nakano {\it et al.}, LEPS Coll., Phys. Rev. Lett. {\bf 91}, 012002 (2003);
V. Barmin {\it et al.}, DIANA Coll., Phys. Atom. Nucl. {\bf 66}, 1715 (2003);
S. Stepanyan {\it et al.}, CLAS Coll.,Phys. Rev. Lett. {\bf 91}, 252001 (2003);
J. Barth {\it et al.}, SAPHIR Coll., hep-ex/0307083.

\bibitem{wilczek}
R. Jaffe and F. Wilczek, Phys. Rev. Lett. {\bf 91}, 232003 (2003).

\bibitem{sasaki}
S. Sasaki, ``{\it `Lattice study of exotic S = +1 baryon}'', 
hep-lat/0310014.

\bibitem{lipkin}
M. Karliner and H. Lipkin, hep-ph/0307243; hep-ph/0307343.

\bibitem{cheung}
K. Cheung, Phys. Rev. {\bf D} in press, hep-ph/0308176.

\bibitem{chiu}
T.W. Chiu and T.H. Hsieh, hep-ph/0404007.

\bibitem{wise}
I.W. Stewart, M.E. Wessling, and M.B. Wise, hep-ph/0402076.

\bibitem{cfrag}
S. Pahhi for for H1 and ZEUS Collaboration, ``{\it Open charm and 
beauty production}'', prepared for Ringberg Workshop on New Trends 
in HERA Physics 2003, Ringberg Castle, Germany, 28 Sep - 3 Oct 2003.

\bibitem{pdg}
Review on ``b-flavored hadrons'' in Particle Data Book,
K. Hagiwara {\it et al.}, Phys. Rev. {\bf D66}, 010001 (2002).

\bibitem{H1-charm}
S. Chekanov for H1 and ZEUS Collaboration, ``{\it Open charm production
in DIS at HERA}'', hep-ex/03090004.

\bibitem{ZEUS-beauty}
ZEUS Collaboration, hep-ex/0405069.

\bibitem{lepew}
The LEP Collaborations, ``{\it A combination of preliminary electroweak
measurements and constraints on the standard model}'', hep-ex/0312023.

\bibitem{cdf-charm}
CDF Collaboration (D. Acosta et al.). Phys. Rev. Lett.{\bf 91} 241804 (2003);

\bibitem{cdf-bottom}
CDF Collaboration (D. Acosta et al.), Phys. Rev. {\bf D65}, 052005 (2002).

\end{thebibliography}
\end{document}